\documentstyle[12pt]{article}
\newcommand{\drm}{\rm d}
\newcommand{\Real}{{\rm Re}}
\newcommand{\Imag}{{\rm Im}}

\newcommand{\Le}{{\cal L}}

\newcommand{\pp}{/\!\!\!p}

\newcommand{\om}{\omega}

\newcommand{\bb}{\begin{equation}}
\newcommand{\ee}{\end{equation}}
\newcommand{\bega}{\begin{eqnarray}}
\newcommand{\ega}{\end{eqnarray}}
\newcommand{\begae}{\begin{eqnarray*}}
\newcommand{\egae}{\end{eqnarray*}}

\newcommand{\ga}{\gamma}

\newcommand{\Sig}{\Sigma}

\newcommand{\Lef}{\Leftrightarrow}

\newcommand{\al}{\alpha}

\newcommand{\C}{I\!\!\!C}

\newcommand{\la}{\lambda}

\newcommand{\Om}{\Omega}

\newcommand{\h}{\hspace*{0.5 cm}}
\newcommand{\hh}{\hspace*{.0 cm}}

\newcommand{\dis}{\displaystyle}

\newcommand{\ov}{\overline}

\newcommand{\vs}{\vspace*}

\newcommand{\Vbf}{\mbox{\boldmath $V$}}
\newcommand{\Abf}{\mbox{\boldmath $A$}}
\newcommand{\sbf}{\mbox{\boldmath $s$}}

\newcommand{\wbf}{\mbox{\boldmath $w$}}
\newcommand{\xbf}{\mbox{\boldmath $x$}}
\newcommand{\ubf}{\mbox{\boldmath $u$}}

\newcommand{\impulse}{\mbox{\boldmath $p$}}

\begin{document}
 
\baselineskip 0.6cm
 
\vspace*{4.cm}

{\bf FIELD THEORY OF THE SPINNING ELECTRON:}
 
{\bf INTERNAL MOTIONS $^{(*)}$}

\footnotetext{$^{(*)}$ Work supported in part by INFN,
CNR, MURST, and by CNPq, FAPESP.}

\vs{.8cm}

\begin{center}
\begin{tabular}{ll}
&\hh Giovanni SALESI$^1$ and Erasmo RECAMI$^2$\\
&\\
&\hh $^1$Dipart.$\!$ di Fisica, Universit\`a Statale di Catania, \\
&\hh 57 Corsitalia, 95129--Catania, Italy.\\
&\hh $^2$Facolt\`a di Ingegneria, Universit\`a Statale di Bergamo, \\
&\hh 24044--Dalmine (BG), Italy;\\
&\hh INFN, Sezione di Milano, Milan, Italy; \ and\\
&\hh Dept. of Applied Math., State University at Campinas,\\
&\hh Campinas, S.P., Brazil.
\end{tabular}
\end{center}
 
\vs{.5mm}
 
\hfill{{\em ``If a spinning particle is not quite a point particle, nor }}

\hfill{{\em a solid three dimensional top, what can it be?"}}

\rightline{Asim O. Barut \qquad \qquad}

\vs{1.2 cm}

{\bf ABSTRACT and INTRODUCTION}\\

\h This paper is dedicated to the memory of Asim O. Barut, who so
much contributed to clarifying very many fundamental issues of physics, 
and whose work constitutes a starting point of these articles.

\h We present here a field theory of the spinning 
electron, by writing down a new equation for the 4-velocity 
field $v^\mu$ (different from that of Dirac theory),
which allows a classically intelligible description of the electron. \ 
Moreover, we
make explicit the noticeable kinematical properties of such 
velocity field (which also result different from the ordinary ones). 
\ At last, we analyze the internal {\em zitterbewegung\/} (zbw)
motions, for both time-like and light-like speeds. \ We adopt in this
paper the ordinary tensorial language. \ Our starting point is the
Barut--Zanghi classical theory for the relativistic electron, which
related spin with zbw.\\

\newpage

{\bf A NEW MOTION EQUATION FOR THE\\ 
SPINNING (FREE) ELECTRON}\\
 
\h Attempts to put forth classical models for the spinning electron
are known since more than seventy
years $^{(1)}$.  In the Barut--Zanghi (BZ) theory,$^{(2)}$ the classical 
electron was actually
characterized, besides by the usual pair of conjugate variables
$(x^\mu, p^\mu)$, by a second pair of conjugate classical {\em spinorial}
variables $(\psi, \ov{\psi})$, representing internal degrees of freedom,
which were functions of the (proper) time $\tau$  measured in the 
electron global
center-of-mass (CM) system; the CM frame (CMF) being the one in which 
$\impulse = 0$ at every instant of time.  \ Barut and Zanghi, then,
introduced a classical lagrangian that in the free case (i.e., when the
{\em external} 
 electromagnetic potential is $A^\mu = 0$) writes $[c=1]$
$$
\Le=\frac{1}{2} i \la (\dot{\ov{\psi}}\psi - \ov{\psi}\dot{\psi}) +
p_\mu(\dot{x}^\mu - \ov{\psi}\ga^\mu \psi) \; , \eqno{\rm (1)}
$$
where $\la$ has the dimension of an action and $\psi$ and
$\ov{\psi}\equiv \psi^{\dag} \ga^0$ are ordinary ${\rm
{\C}}^4$--bispinors, the dot meaning derivation with respect to
$\tau$. \ The four Euler--Lagrange equations, with $-\la=\hbar=1$, 
yield the following motion equations: 
 
\
 
$\hfill{\dis\left\{\begin{array}{l}
\dot{\psi}+ i p_\mu \ga^\mu \psi=0\\

\dot{x}^\mu= \ov{\psi}\ga^\mu \psi\\

\dot{p}^{\mu}=0 \; ,
\end{array}\right.}
\hfill{\dis\begin{array}{r}
(2{\rm a}) \\ (2{\rm b}) \\ (2{\rm c}) \end{array}}$
 
\
 
besides the hermitian adjoint of eq.(2a), holding for $\ov{\psi}$. \ 
From eq.(1) one can also see that
$$
H \equiv p_{\mu} v^{\mu} = p_{\mu} \ov{\psi} \ga^{\mu} \psi
\eqno{\rm (3)}
$$
is a constant of the motion [and precisely is the energy in the
CMF].$^{(2-4)}$  \ Since $H$ is the BZ hamiltonian in the CMF, we
can suitably set \ $H = m$, \ quantity $m$ being the particle
rest-mass. \ \ The general solution of the equations of motion (2)
can be shown to be:
$$
\psi(\tau)=[\cos (m\tau)- i \frac{p_\mu \ga^\mu}{m}\sin (m\tau)]\psi(0) \  ,
\eqno{\rm(4 a)} $$

$$
\ov{\psi}(\tau)=\ov{\psi}(0)[\cos (m\tau)+ i \frac{p_\mu \ga^\mu}{m}
\sin (m\tau)] \  ,  \eqno{\rm(4 b)} $$
 
with $p^\mu=$ constant; \ $p^2=m^2$; \ and finally:   %%%  ??????????????????
$$
\dot{x}^\mu\equiv v^\mu=\frac{p^\mu}{m}+[\dot{x}^\mu(0)-\frac{p^\mu}{m}]
\cos(2m\tau)+\frac{\ddot{x}^\mu}{2m}(0)\sin (2m \tau) \ .
\eqno{\rm(4 c)} $$
 
This general solution exhibits a classical analogue of the phenomenon known 
as zitterbewegung: in fact, the velocity $v^\mu$ contains the
(expected) term $p^\mu/m$ plus a term describing an oscillating
motion with the characteristic zbw frequency $\om=2m$. \ The velocity
of the CM will be given by $p^\mu/m$. \  \ Let us explicitly observe that the  
general solution (4c)
represents a helical motion in the ordinary 3-space of a ``constituent"
${\cal Q}$: a result that has been met
also by means of other, alternative approaches.$^{(5,6)}$ 

\newpage

\h Before studying the time evolution of our electron, we want to write down
its motion equation in a ``kinematical'' form, suitable a priori for
describing a point-like object; i.e., at variance with eqs.(2), expressed
not in terms of $\psi$ and $\ov{\psi}$, but on the contrary in terms of 
quantities related to the particle trajectory (such as $p^\mu$ and 
$v^\mu$). \ To this aim, we can introduce the
spin variables, and adopt the set of dynamical variables
\[
x^\mu \, , \ p^\mu \, ; \ v^\mu \, , S^{\mu \nu} \; ,
\]
where
$$
S^{\mu \nu} \equiv {i \over 4} \, \ov{\psi} [\ga^\mu, \ga^\nu] \psi \; ;
\eqno{\rm(5a)} $$
 
then, we get the following motion equations:
$$
\dot{p}^\mu=0 \ ; \ \ v^\mu=\dot{x}^\mu \ ; \ \ \dot{v}^\mu=4 S^{\mu
\nu}p_{\nu} \ ; \ \ \dot{S}^{\mu \nu}= v^\nu p^\mu - v^\mu p^\nu \ .
\eqno{\rm(5b)}  $$
 
[By varying the action corresponding to $\Le$, one finds as
generator of space-time rotations the conserved quantity \ $J^{\mu
\nu}=L^{\mu\nu}+ S^{\mu\nu}$, \ where \ $L^{\mu\nu} \equiv x^\mu p^\nu - 
x^\nu p^\mu$ \ is the orbital
angular momentum tensor, and $S^{\mu\nu}$ is just the particle spin tensor: 
 \ so that \ 
${\dot{J}}^{\mu\nu}=0$ \ implies \ $\dot{L}^{\mu\nu}=-\dot{S}^{\mu\nu}$].
\hfill\break
\h By deriving the third one, and using the first one, of eqs.(5b),
we obtain
$$
\ddot{v}^\mu = 4\dot{S}^{\mu\nu} p_\nu \ ;
\eqno{\rm (6)} $$ 
by substituting now the fourth one of eqs.(5b) into eq.(6), and imposing
the previous constraints $p_\mu p^\mu=m^2$ \ and \ $p_\mu v^\mu=m$, we 
end with the time evolution$^{(3)}$ of the {\em field four-velocity}:
$$
v^\mu= \frac{p^\mu}{m}-\frac{\ddot{v}^\mu}{4 m^2} \; , \eqno{\rm(7)} 
$$
such {\em a new motion equation} corresponding to the whole system of eqs.(2).
 \ Let us recall, for comparison, that the analogous
equation {\em for the standard Dirac case\/}:$^{(1)}$ 
$$
v^\mu = {\dis\frac{p^\mu}{m}- \frac{i}{2m}} \dot{v}^\mu 
\eqno{\rm (7')} 
$$
was totally devoid of a
classical, intuitive meaning, because of the known appearance of an imaginary 
unit $i$ in front of the acceleration (connected with the well-known fact that
the position operator is not hermitian therein).
 
\h Let us observe that, by differentiating the relation
$p_\mu v^\mu = m = {\rm constant}$, one immediately finds that the (internal)
acceleration $\dot{v}^\mu\equiv \ddot{x}^\mu$ is orthogonal to the electron
impulse $p^\mu$, since $p_\mu \dot{v}^\mu=0$ at any instant. To conclude, let 
us stress that, while the Dirac electron has
no classically meaningful internal structure, our electron on the contrary
(an {\em extended--type} particle) does possess an internal
structure, and internal motions, which are all endowed with a 
``realistic" meaning, from both the geometrical and kinematical points of
view: as we are going to see in the next section.\\

\newpage

{\bf SPIN AND INTERNAL KINEMATICS}\\                           

\h We wish first of all to make explicit the kinematical definition of
$v^\mu$, {\em rather different from the ordinary one} valid for scalar 
particles.$^{(7)}$. \ In fact, from
the very definition of $v^\mu$, we get
\[
v^\mu\equiv \drm x^\mu/ \drm \tau \equiv (\drm t/ \drm \tau; 
\drm \xbf / \drm \tau)
\equiv ({\frac{\drm t}{\drm \tau}}; \frac{\drm \xbf}{\drm t} \; 
\frac{\drm t}{\drm \tau})
\]
$$
=(1/ \sqrt{1- \wbf^2}; \;\; \ubf / \sqrt{1- \wbf^2}) \; , \;\;\;\;\; \quad
[\ubf \equiv \drm \xbf / \drm t] \eqno{\rm(8)}
$$
where $\wbf = \impulse/ m$ is the velocity of the CM in the chosen
reference frame (i.e., in the frame in which quantities $x^\mu$ are 
measured). \ Below, it will be convenient to choose as reference frame the
CMF  (even if quantities as $v^2 \equiv v_\mu v^\mu$
are frame invariant);  so that
$$
v^\mu_{\rm CM}= V^\mu \equiv (1; \Vbf) \ , \eqno{\rm(9)}
$$
wherefrom one deduces for the speed $|\Vbf|$
of the internal motion (i.e., for the zbw speed) the {\ em new} conditions:
\[
0 < V^2 (\tau) < 1 \;\;\; \Lef \;\;\; 0 < \Vbf^2 (\tau) < 1 \;\;\;\;\;\;
\quad\mbox{(time-like)}
\]
$$
V^2(\tau)=0 \quad \Lef \quad \Vbf^2 (\tau) = 1 \ \ \;\;\;\;\; \ \ \ \ \ \ \ \ \ 
 \quad\mbox{(light-like)}
\eqno{\rm (10)} $$ 
\[
V^2 (\tau) < 0 \quad \Lef \quad \Vbf^2 (\tau) > 1 \ \ \;\;\;\;\; \ \ \
\ \ \ \ \ \ \ \quad\mbox{(space-like)} \ ,
\]
where $V^2 = v^2$. \  Notice that, in general, the value of $V^2$ does vary
with $\tau$; except in special cases (e.g., the case of polarized particles: 
as we shall see).
 Coming back to the expression of the 4-velocity, eq.(4c), it is possible
 after some algebra to recast this equation in a ``spinorial'' form, i.e.,
 to write it as a function of the initial spinor $\psi(0)$:
$$
v^\mu= p^\mu /m + E^\mu \cos (2m \tau)+ H^\mu\sin (2m \tau) \; ,
\eqno{\rm (11)} 
$$
where [$\al^{\mu}\equiv\ga^0\ga^{\mu}$]
$$
E^\mu= \frac{1}{2}\ov{\psi}(0)[\frac{\pp}{m}\;,\; \al^{\mu}]\psi(0)
\eqno{\rm (12a)}$$                                                      
$$                           
H^\mu= \frac{i}{2}\ov{\psi}(0)(\al^{\mu} - \frac{\pp}{m}\al^{\mu}\frac{\pp}{m})
\psi(0) \; . \eqno{\rm (12b)}
$$                                                                                                                                             
In the chosen CM frame, eqs.(12) read:
$$E^\mu= \ov{\psi}(0)\ga^{\mu}\psi(0) - \frac{p^\mu}{m} \eqno{\rm (13a)}
$$                                                                                                          
$$H^\mu = i\ov{\psi}(0)(\al^{\mu} - g^{0\mu})\psi(0) \; , \eqno{\rm (13b)}
$$                                                      
where $g^{\mu \nu}$ is the metric tensor. \ Bearing in mind that (in 
the CMF) it holds  $v^0=1$ \ [cf. eq.(9)], and therefore $\ov{\psi} \ga^0 \psi=1$ 
(which, incidentally,
implies the normalization $\psi^{\dag} \psi=1$ in the CMF), \ one obtains
$$
E^\mu =
(0; \; \ov{\psi} (0) \vec{\ga} \psi (0)) \ 
\eqno{\rm (14a)} 
$$

$$
H^\mu =(0; \;
i \ov{\psi}(0) \vec{\al} \psi(0)) \; .
\eqno{\rm (14b)} 
$$
By eq.(4), for $V^2$ we can write:
$$
V^2 = 1+ E^2 \cos^2 (2m \tau)+ H^2 \sin^2(2m \tau)+
2 E_\mu H^\mu \sin (2m \tau) \cos (2m \tau) \; .
\eqno{\rm(15)}.
$$
\h Now, let us single out the solutions $\psi $ of eq.(2)
corresponding to {\em constant} $V^2$ and $A^2$, where $A^{\mu}
\equiv \drm V^\mu / \drm \tau \equiv (0; \Abf)$, quantity $V^\mu
\equiv (1; \Vbf)$ being the zbw velocity. In the present frame,
therefore, we shall suppose quantities
\[
V^2= 1- \Vbf^2 \ \ ; \ \ A^2= -\Abf^2
\]
to be constant in time:
$$
V^2 \ = \ {\rm constant} \;; \ \ \ \  A^2 \ = \ {\rm constant} \; ,
\eqno{\rm(16)}
$$
so that $\Vbf^2$ and $\Abf^2$ are constant in time too. (Let us
recall that we are dealing with the internal motion only, in the CMF;
thus, our results are independent of the global 3-impulse $\impulse$
and hold both in the relativistic and in the non-relativistic case).
\ Requirements (16), inserted into eq.(15), yield the following
interesting constraints:$^{(7)}$ 

\

$\hfill{\dis\left\{\begin{array}{l}
E^2= H^2\\
E_\mu H^\mu=0 \; .\\
\end{array}\right.}
\hfill{\dis\begin{array}{r}
(17{\rm a}) \\ (17{\rm b}) \end{array}}$
 
\
 
\h Constraints (17) are necessary and sufficient (initial) conditions
to get a circular {\em uniform} motion (the only finite, uniform
motion possible in the CMF). \ Since
both $E$ and $H$ do not depend on $\tau$, also eqs.(17) hold
at any time. \ In the euclidean 3-dimensional space, 
and at any time, constraints (17) may read:

\

$\hfill{\dis\left\{\begin{array}{l}
\Abf^2= 4 m^2 \Vbf^2\\
\Vbf \cdot \Abf = 0\\
\end{array}\right.}
\hfill{\dis\begin{array}{r}
(18{\rm a}) \\ (18{\rm b}) \end{array}}$
 
\
 
which explicitly correspond to a uniform circular motion with radius
$$
R= |\Vbf|/ 2 m \ .
\eqno{\rm (19)} $$
 
Quantity $R$ is the ``zitterbewegung radius"; the zbw frequency
was already found to be $\Om= 2m$. \  By means of eqs.(14),
conditions (17) or (18) can be written in spinorial form (still for any
time instant $\tau$) as follows:
 
\

$\hfill{\dis\left\{\begin{array}{l}
(\ov{\psi} \vec{\ga} \psi)^2= -(\ov{\psi} \vec{\al} \psi)^2\\
(\ov{\psi}\vec{\ga}\psi) \cdot (\ov{\psi} \vec{\al}\psi)=0 \; .
\end{array}\right.}
\hfill{\dis\begin{array}{r}
(20{\rm a}) \\ (20{\rm b}) \end{array}}$

\
 
\h At this point, let us show that this classical
uniform circular motion, around the  $z$-axis (which in the
CMF can be chosen arbitrarily, while in a generic
frame is parallel to the global three-impulse $\impulse$, as we shall see  
below), does just correspond to the case of {\em polarized} particles
with $s_z = \pm {1 \over 2}$. \ It may be interesting to notice
that in this case the {\em classical} requirements (17) or (18) ---namely,
the uniform motion conditions--- play the role of the ordinary
{\em quantization} conditions \ $s_z = \pm {1 \over 2}$.                                      

\h It is straightforward to realize also that the most general
spinors $\psi(0)$ satisfying the conditions
$$
s_x = s_y=0
\eqno{\rm(21a)} $$
$$
s_z = {\frac{1}{2}\ov{\psi}(0)\Sig_z \psi(0)} = \pm {1 \over 2}
\eqno{\rm(21b)} $$
 
($\vec{\Sig}$ being the spin operator) possess in the standard 
representation the form
$$
\psi_{(+)}^{\rm T} (0)= ( \ a \ 0 \ | \ 0 \ d \ )
\eqno{\rm(22a)} $$
$$
\psi_{(-)}^{\rm T} (0)= ( \ 0 \ b \ | \ c \ 0 \ ) \; , 
\eqno{\rm(22b)} $$

and obey in the CMF the normalization constraint $\psi^{\dag}
\psi=1$. \ [It could be easily shown that, for generic initial
conditions, it is always \ $-{1 \over 2} \leq s_z \leq {1 \over 2}$].
\ Notice that the set of our spinors $\psi_{\pm}$ include the Dirac
spinors, but is an ensemble larger than Dirac's. \ In
eqs.(22) we separated the first two from the second two components,
bearing in mind that in the standard Dirac theory (and in the CMF)
they correspond to the positive and negative frequencies,
respectively. \ With regard to this point, let us observe that the
``negative-frequency"
components $c$ and $d$ do {\em not} vanish at the non-relativistic
limit (since, let us repeat, in the CMF it is $\impulse = 0$); but
the field hamiltonian $H$ is {\em nevertheless} positive and equal to
$m$, as already stressed. \ Now, from relation (22a) we are able  to
deduce that (with $* \equiv$ complex conjugation):
\begae
< \vec{\ga} > &\equiv& \ov{\psi} \vec{\ga} \psi= 2( \Real [a^* d], + \Imag
[a^* d], 0)\\
< \vec{\al} > &\equiv& \ov{\psi} \vec{\al} \psi= 2i( \Imag [a^* d], - \Real
[a^* d], 0)
\egae
and analogously, from eq.(22b), that
\begae
< \vec{\ga} > &\equiv& \ov{\psi} \vec{\ga} \psi= 2( \Real [b^* c], - \Imag
[b^* c], 0)\\
< \vec{\al} > &\equiv& \ov{\psi} \vec{\al} \psi= 2i( \Imag [b^* c], + \Real
[b^* c], 0) \; ,
\egae
which just imply relations (20):
 
\[
\left\{\begin{array}{l}
< \vec{\ga} >^2 = - < \vec{\al} >^2\\
\\
< \vec{\ga} > \cdot < \vec{\al} > =0 \ .
\end{array}\right.
\]
 
\h In conclusion, the (circular) polarization conditions, eqs.(21), do
imply the internal zbw motion to be uniform and circular ($V^2=$
constant; $A^2=$ constant); \ equations (21), in other words, imply
simultaneously that $s_z$ be conserved and quantized.$^{(7)}$.
 
\h When passing from the CMF to a generic frame,
eqs.(21) transform into
$$
\la \equiv {1 \over 2} \ov{\psi} (x) \frac{\vec{\Sig} \cdot
\impulse}{|\impulse|} \psi (x) \; = \; \pm {1 \over 2} \; = \; {\rm
constant} \  .\eqno{\rm(23)}
$$
 
Therefore, to get a uniform motion around the
$\impulse$-direction [cf. eq.(4c)], we have to require
that the field helicity $\la$ be constant (in space and in time), and
quantized in the ordinary way: \ $\la = {1 \over 2}$.

\h It may be interesting also to calculate $|\Vbf|$ as a function
of the spinor components $a$ and $d$. \ With reference to eq.(22a), since
$\psi^{\dag} \psi \equiv |a|^2 + |d|^2=1$, we obtain (for the $s_z= + 
{1 \over 2}$
case):
$$
\Vbf^2 \equiv < \vec{\ga} >^2= 4| a^* d|^2=4 |a|^2 \; (1-|a|^2)
\eqno{\rm(24a)} $$

$$
\Abf^2 \equiv (2 i m < \vec{\al} >)^2= 4m^2 \Vbf^2=16 m^2 |a|^2
 \; (1-|a|^2)\ ,
\eqno{\rm(24b)} $$
and therefore the normalization value (valid now in any frame, at any time): 
$$
\ov{\psi} \psi = \sqrt{1 - \Vbf^2} \ ,
\eqno{\rm(24c)} $$
showing that to the same speed and acceleration there correspond two spinors
$\psi(0)$, related by an interchange of $a$ and $d$. \ From eq.(24a)
we derive also that, as $0 \leq |a| \leq 1$, it is:
$$
0 \leq \Vbf^2 \leq 1 \; ; \ \ \ 0 \leq \ov{\psi} \psi \leq 1 \; .
\eqno{\rm(24d)} $$
 
Correspondingly, from eq.(19c) we would obtain for the zbw radius that \ 
$0 \leq R \leq {1 \over 2} m$.

\h The second of eqs.(24d) is a {\em new}, rather interesting 
(normalization) boundary 
condition. From eq.(24c) one can easily see that: \ (i) for $\Vbf^{2} = 0 $ 
(no zbw) we
have $\ov{\psi} \psi = 1 $ and $\psi$ is a ``Dirac spinor''; \ (ii) 
for $\Vbf^{2} = 1$
(light-like zbw) we have $\ov{\psi} \psi = 0 $ and $\psi$ is a ``Majorana'' 
 spinor''; \ (iii) for $ 0 < \Vbf^2 <1 $ we meet, instead, spinors with 
characteristics  ``intermediate'' between the Dirac and Majorana 
ones.
  
\h The ``Dirac" case, corresponding to $\Vbf^2=
\Abf^2 = 0$, \ i.e., to {\em no} zbw internal motion, is
trivially represented (apart from phase factors) by the spinors:

$$
\psi^{\rm T} (0) \equiv (1 \ \ 0 \ | \ 0 \ \ 0) \eqno{\rm(25)}
$$
and (interchanging $a$ and $d$):
$$
\psi^{\rm T}(0) \equiv (0 \ \ 0 \ | \ 0 \ \ 1) \ .
\eqno{\rm(25')} $$
 
This is the unique case (together with the analogous one for $s_z = 
-\frac{1}{2}$) in which the zbw  disappears, while the
field spin is still present! In fact, even in terms of
eqs.(25)--(25') one still gets that \ $\frac{1}{2} \ov{\psi} \Sig_z
\psi= +\frac{1}{2}$.\hfill\break

\h Since we have been discussing a classical theory of the relativistic electron,
  let us finally notice that even the
well-known change in sign of the fermion wave function, under
360$^{\rm o}$-rotations around the $z$-axis, gets in our theory a 
natural classical interpretation. \ In fact, a 360$^{\rm o}$-rotation
of the coordinate frame around the $z$-axis (passive point of view) is indeed
equivalent to a 360$^{\rm o}$-rotation of the constituent $\cal Q$ around
the $z$-axis (active point of view). \ On the other hand, as a consequence of 
the latter transformation, the zbw angle $2 m\tau$ does suffer a  
variation of 360$^{\rm o}$, \ the
proper time $\tau$ does increase of a zbw period $T=\pi/m$, \ and the
pointlike constituent does describe a complete circular orbit around the
$z$-axis. \ It appears then obvious that, since 
the period $T= 2\pi / m$ of spinor $\psi(\tau)$ in
eq.(4c) is {\em twice} as large as the zbw orbital period, the wave
function of eq.(4c) does suffer a phase--variation of
180$^{\rm o}$ only, and then does change its sign: as it occurs in the standard
theory.\\

\

{\bf SPECIAL CASES: LIGHT-LIKE MOTIONS\\ 
AND LINEAR MOTIONS}\\

\h Let us first fix our attention on the special
case of {\em light-like} motions.$^{(7,6)}$ \ The spinor fields $\psi(0)$,
corresponding to $V^2=0; \Vbf^2=1$, \  are given by eqs.(22) with \ 
$|a|=|d| \quad \mbox{for the}\quad s_z= +{1 \over 2} \quad\mbox{case}$, \
or \ $|b|=|c| \quad \mbox{for the}\quad s_z = -{1 \over 2} \quad\mbox{case}$; \
as it follows from eqs.(24) for $s_z = +{1 \over 2}$, as well as  from 
the analogous equations
$$
\Vbf^2=4|b^* c|=4|b|^2 (1-|b|^2)
\eqno{\rm (26a)} $$
$$
\Abf^2=4 m^2 \Vbf^2 = 16 m^2 |b|^2 (1-|b|^2) \; ,
\eqno{\rm (26b)} $$
 
for the case $s_z = -{1 \over 2}$. \ It can be easily seen that a 
difference in the phase
factors of $a$ and $d$ (or of $b$ and $c$, respectively) does {\em not} 
change the motion kinematics, nor its rotation direction; \
but it merely shifts the zbw phase angle at $\tau = 0$. \ Thus, one is 
entitled to choose {\em purely real} spinor components (as we did above).  
As a consequence, the {\em simplest} spinors may be written:
$$
\psi_{(+)}^{\rm T} = {\dis\frac{1}{\sqrt{2}}} (1 \ \ 0 \ | \ 0 \ \ 1)
\eqno{\rm (27a)} $$
$$
\psi_{(-)}^{\rm T} = {\dis\frac{1}{\sqrt{2}}} (0 \ \ 1 \ | \ 1 \ \ 0)\
;
\eqno{\rm (27b)} $$
 and then
\[
< \vec{\ga} >_{(+)} = (1, 0, 0) \ ; \ < \vec{\al} >_{(+)} = (0, -i, 0)
\]
\[
< \vec{\ga} >_{(-)} = (1, 0, 0) \ ; \ < \vec{\al} >_{(-)} = (0, i, 0)
\]
which, inserted into eqs.(14), yield  
\[
{E^\mu}_{(+)} =(0; 1, 0, 0) \; ; \qquad {H^\mu}_{(+)} = (0; 0,1,0) \; .
\]
\[
{E^\mu}_{(-)} =(0; 1, 0, 0) \; ; \qquad {H^\mu}_{(-)} = (0; 0,-1,0) \; .
\]

Because of eq.(11), we meet now
 for $s_z = +{1 \over 2}$ an {\em anti-clockwise\/} internal motion, 
with respect to the chosen $z$-axis:
$$
V_x= \cos (2 m \tau); \quad V_y=\sin (2 m \tau); \quad V_z=0 \ ;
\eqno{\rm(28)} $$

and a {\em chockwise} internal motion for $s_z = -{1 \over 2}$:
$$
V_x=\cos (2m \tau);\quad V_y=-\sin(2 m\tau); \quad V_z=0 \; .
\eqno{\rm(29)} $$
 
\h Let us explicitly observe that spinor (27a), associated with
$s_z = +{1 \over 2}$ (i.e., with an anti-clockwise internal rotation), gets 
contributions of equal 
magnitude from the positive--frequency spin-up component
and from the negative--frequency spin-down component: in full
agreement with our ``reinterpretation" in terms of particles and
antiparticles, given in refs.$^{(8)}$. \  Analogously, spinor (27b), 
associated with
$s_z = -{1 \over 2}$ (i.e., with a clockwise internal rotation), gets
contributions of equal magnitude from the positive-frequency spin-down
component and the negative-frequency spin-up component.$^{(8)}$
 
\h Let us observe also that, having recourse to the light-like
solutions, one is actually entitled to regard the electron spin as
totally arising from the zbw motion, since the {\em intrinsic} term
$\Delta^{\mu \nu}$ entering the BZ theory$^{(2)}$ does {\em vanish}
when $v^{\mu}$ tends to $c$.

\h As we have seen above [cf. eq.(23)], in a {\em generic} reference frame
the polarized states are characterized by a helical uniform motion
around the $\impulse$-direction; therefore, the 
$\la = +{1 \over 2} \;\;\; [\la =
-{1 \over 2}]$ spinor will
correspond  to an anti-clockwise [a clockwise] helical motion
with respect to the $\impulse$-direction.
 
\h Going back to the CMF, we have to remark that  
eq.(19) yields in this case for the zbw radius $R$ the traditional result:
$$
R=\frac{|\Vbf|}{2 m} \equiv \frac{1}{2m} \equiv \frac{\la}{2} \; ,
\eqno{\rm(30)} $$
where $\la$ is the Compton wave-length. \ Of course, \ $R = {1 \over 2} m$ \
represents the {\em maximum} size (in the CMF) of the electron, among all the
uniform motion ($A^2=$const.; \ $V^2=$const.) solutions.  The minimum,
$R=0$, corresponding to the limiting Dirac case with no zbw ($V=A=0$), represented 
by eqs.(25), (25'): so that the Dirac free
electron is a pointlike, extensionless object.\\

\h Before concluding this Section, let us shortly consider what happens
when {\em releasing} the conditions (22)--(25)
(and therefore abandoning the assumption of circular uniform motion), 
 so to obtain an internal oscillating motion along a constant straight line. 
For instance, one may choose either
$$
\psi^{\rm T}(0) \equiv \frac{1}{\sqrt{2}} (1 \ \ 0 \ | \ 1 \ \ 0) \; ,
\eqno{\rm(31)} $$
 
or \ \ $\psi^{\rm T}(0) \equiv \frac{1}{\sqrt{2}} (1 \ \ 0 \ | \ i \ \
0)$, \ or \ \ $\psi^{\rm T}(0) \equiv \frac{1}{2} (1 \ \ -1 \ | \ -1 \ \ 1)$,
 \ or \ \ $\psi^{\rm T}(0) \equiv \frac{1}{\sqrt{2}} (0 \ \ 1 \ | \ 0 \ \ 1)$,
 \ and so on.
 
In case (31), for example, one actually gets
$$
< \vec{\ga} > \equiv (0,0,1) \ ; \ < \vec{\al} > \equiv (0,0,0)$$
which, inserted into eqs.(14), yield
$$
E^\mu = (0;0,0,1) \ ; \ H^\mu = (0;0,0,0).
$$
Therefore, because of eq.(21a), we have now a {\em linear,
oscillating} motion [for which equations (22), (23), (24) and (25) do {\em
not} hold: here $V^2(\tau)$ does vary from 0 to 1!] along the $z$-axis:
\[
V_x(\tau)=0; \quad V_y(\tau)=0; \quad V_z(\tau)=\cos(2 m \tau) \ .
\]
 
All the spinors written above could describe an unpolarized, mixed state, 
since it holds
\[
\sbf \equiv \frac{1}{2} \ov{\psi} \vec{\Sig}\psi = (0, 0, 0) \ ,
\]
in agreement with the existence of a linear oscillating motion. \h 
Furthermore for such new spinors it holds \ $\ov{\psi} \psi = 
\ov{\psi}\ga^5\psi = 0$, \ but \  
$\ov{\psi}\ga^5\ga^{\mu}\psi \neq 0$ and $\ov{\psi}S^{\mu\nu}\psi \neq 0$.
This {\em new} class of spinors has been very recently proposed and 
extensively studied
by Lounesto,$^{(9)}$  by employing a new concept, called ``boomerang'', 
within the
framework of Clifford algebras. A physical realization of those new
spinors$^{(9)}$ seems now to be provided by our electron, in the 
present case. \\

\

{\bf ACKNOWLEDGEMENTS}\\

The authors
are grateful to  J.P. Dowling for having extended to them the permission to
contribute to this Volume of Proceedings in memory of Professor Barut. \ They
acknowledge continuous, stimulating discussions with H.E.Hern\'andez,
M. Pav\v{s}i\v{c}, S. Sambataro, D. Wisnivesky, J. Vaz and particularly W.A. 
Rodrigues Jr. \  
Thanks for useful discussions and kind collaboration are also due 
G. Andronico, G.G.N. Angilella, M. Baldo, M. Borrometi, A. Buonasera, 
A. Bugini, 
F. Catara, A. Del Popolo, C.~Dipietro, M. Di Toro, G. Giuffrida, 
A.A. Logunov, J. Keller, C. Kiihl, G.D. Maccarrone, J.E. Maiorino,
R. Maltese, G. Marchesini, R. Milana, R.L. Monaco, E.C. de Oliveira, 
M. Pignanelli, P.I. Pronin, G.M. Prosperi, 
M. Sambataro, J.P. dos Santos, P.A. Saponov, G.A. Sardanashvily, Q.A.G. 
Souza, E. Tonti, P. Tucci, R, Turrisi, M.T. Vasconselos and J.R. Zeni.\\

\


\begin{thebibliography}{8}

\bibitem {1} A.H. Compton: Phys. Rev. {\bf 14} (1919) 20, 247; \ E.
Schr\"odinger: Sitzunger. Preuss. Akad. Wiss. Phys. Math. Kl. {\bf 24} (1930) 
418. \ See also P.A.M. Dirac: {\em The Principles of Quantum Mechanics},  
$4^{\rm th}$ edition (Claredon; Oxford, 1958), p.262; \
J. Frenkel: Z. Phys. {\bf 37} (1926)
243; \ M. Mathisson: Acta Phys. Pol. {\bf 6} (1937) 163;
H. H\"{o}nl and A. Papapetrou: Z. Phys. {\bf 112}
(1939) 512; {\bf 116} (1940) 153; \ M.J. Bhabha
and H.C. Corben: Proc. Roy. Soc. (London) A{\bf 178} (1941) 273; 
\ K. Huang: Am. J. Phys. {\bf 20}
(1952) 479; \ H. H\"{o}nl: Ergeb. Exacten Naturwiss. {\bf 26} (1952) 29; 
\ A. Proca: J. Phys. Radium {\bf 15} (1954) 5; 
\ M. Bunge: Nuovo Cimento {\bf 1} (1955) 977; \
\ F. Gursey: Nuovo Cimento {\bf 5} (1957) 784; \
W.H. Bostick: ``Hydromagnetic model of an elementary particle", in 
{\em Gravity Res. Found. Essay Contest} (1958 and 1961), and refs. therein;  
 \ H.C. Corben: Phys. Rev. {\bf 121} (1961) 1833; \ 
T.F. Jordan and M. Mukunda: Phys. Rev. {\bf 132} (1963) 1842; \
B. Liebowitz: Nuovo Cimento A{\bf 63} (1969) 1235; \ 
H. Jehle: Phys. Rev. D{\bf 3} (1971) 306; 
\ F. Riewe: Lett. Nuovo Cim. {\bf 1} (1971) 807;
\ G.A. Perkins: Found. Phys. {\bf 6} (1976) 237; 
\ D. Gutkowski, 
M. Moles and J.P. Vigier: Nuovo Cim. B{\bf 39} (1977) 193; \
A.O. Barut: Z. Naturforsch. A{\bf 33}
(1978) 993; \ J.A. Lock: Am. J. Phys. {\bf 47} (1979) 797; \  
M. Pauri: in {\em Lecture Notes in Physics, vol.135\/} (Springer; Berlin,
1980), p.615; \ 
J. Maddox: ``Where Zitterbewegung may lead", Nature {\bf 325} (1987) 306; \
M. H. McGregor: {\em
The enigmatic electron} (Kluwer; Dordrecht, 1992);
\ W.A. Rodrigues, J. Vaz
and E. Recami: Found. Phys. {\bf 23} (1993) 459.


\bibitem {2} A.O. Barut and N. Zanghi: Phys. Rev. Lett. {\bf 52}
(1984) 2009. \ See also A.O. Barut and A.J. Bracken: Phys. Rev. D{\bf 23}
(1981) 2454; \ D{\bf 24} (1981) 3333; \ A.O. Barut and M. Pavsic: 
Class. Quantum Grav. {\bf 4} (1987) L131; \ A.O. Barut: Phys. Lett. B{\bf 237}
(1990) 436.

\bibitem {3} E. Recami and G. Salesi: ``Field theory of the electron: Spin 
and zitterbewegung", in {\em Particles, Gravity and Space-Time}, ed. by P.I. 
Pronin and G.A. Sardanashvily (World Scient.; Singapore, 1996), pp.345-368.

\bibitem {4} M. Pavsic, E. Recami, W.A. Rodrigues, G.D.
Maccarrone, F. Raciti and G. Salesi: Phys. Lett. B{\bf 318} (1993) 481.

\bibitem {5} M. Pavsic: Phys. Lett. B{\bf 205} (1988) 231; \  B{\bf 221}
(1989) 264; \ Class. Quant. Grav. {\bf 7} (1990) L187.

\bibitem {6} A.O. Barut and M. Pavsic: Phys. Lett. B{\bf 216} (1989)
297; \ F.A. Ikemori: Phys. Lett. B{\bf 199} (1987) 239.\ See also D. Hestenes:
 Found. Phys. {\bf 20} (1990)
1213; \ S. Gull, A. Lasenby and C. Doran: ``Electron paths, tunneling
and diffraction in the space-time algebra", to appear in Found.
Phys. (1993); \ D. Hestenes and A. Weingartshofer (eds.): {\em The
electron} (Kluwer; Dordrecht, 1971), in particular the
contributions by H. Kr\"{u}ger, by R. Boudet, and by S. Gull; \
A. Campolattaro: Int. J. Theor. Phys. {\bf 29} (1990) 141; \
D. Hestenes: Found. Phys. {\bf 15} (1985) 63.

\bibitem {7} W.A. Rodrigues Jr., J. Vaz, E. Recami and G. Salesi:
Phys. Lett. B{\bf 318} (1993) 623. 

\bibitem {8} For the physical interpretation of the negative frequency
waves, without any recourse to a ``Dirac sea", see e.g. E. Recami:
Found. Phys. {\bf 8} (1978) 329; \ E. Recami and W.A. Rodrigues:
Found. Phys. {\bf 12} (1982) 709; {\bf 13} (1983) 533; \ M. Pavsic and
E. Recami: Lett. Nuovo Cim. {\bf 34} (1982) 357. \ See also R. Mignani and E.
Recami: Lett. Nuovo Cim. {\bf 18} (1977) 5; \ A. Garuccio {\em et al.\/}: Lett.
Nuovo Cim. {\bf 27} (1980) 60.

\bibitem {9} P. Lounesto: ``Clifford algebras, relativity and quantum
mechanics'', \ in {\em Gravitation: The Space-Time Structure --- Proceedings
of Silarg-VIII}, ed. by W.A. Rodrigues et al. (World Scient.; Singapore,
1994).


\end{thebibliography}
\end{document}